\documentclass[traditabstract]{aa} 
\usepackage{graphicx}
\usepackage{txfonts}
\usepackage{verbatim}
\usepackage{amssymb}
\usepackage{natbib}
\bibpunct{(}{)}{;}{a}{}{,}
\newcommand{\wat}{H$_2$O}
\newcommand{\met}{CH$_3$OH}

\newcommand{\kms}{km~s$^{-1}$}

\newcommand{\ms}{$M_{\odot}$}
\newcommand{\ls}{$L_{\odot}$}

\newcommand{\msyr}{$M_{\odot}$~yr$^{-1}$}
\newcommand{\pas}{$\rlap{.}^{\prime\prime}$}
\newcommand{\degree}{$^{\circ}$}

\begin{document}
%
  \title{Infall and outflow within 400 AU from a high-mass protostar}
   \subtitle{3-D velocity fields from methanol and water masers in AFLG~5142}

  \author{ C. Goddi \inst{1}
\and L. Moscadelli \inst{2}
\and A. Sanna  \inst{3}
}

   \institute{ European Southern Observatory, Karl-Schwarzschild-Strasse 2,
D-85748 Garching  bei M\"{u}nchen, Germany \and INAF, Osservatorio Astrofisico di Arcetri, Largo E. Fermi 5, 50125 Firenze, Italy
    \and  	
Max-Planck-Institut  f\"{u}r Radioastronomie, Auf dem H\"{u}gel 69, 53121 Bonn, Germany
}

\abstract{
Observational signatures of infalling envelopes and outflowing material in early 
stages of protostellar evolution, and at small radii from the protostar, are essential to progress in the understanding
of the mass-accretion process in star formation.
In this letter, we report a detailed study of the accretion and outflow structure around a protostar in the well-known high-mass star-forming region AFGL 5142.
We focus on the mm source MM--1, which  exhibits hot-core chemistry, radio continuum emission, and  strong water (H$_2$O) and methanol (CH$_3$OH) masers. 
 Remarkably, our Very Long Baseline Interferometry (VLBI) observations of molecular masers over six years
provided us with the 3-D velocity field of circumstellar molecular gas with a resolution of 0.001--0.005\arcsec~and at radii $<$0\pas23 (or 400 AU) from  the protostar. 
In particular, our measurements of CH$_3$OH maser emission enabled, for the first time, a direct measurement of infall of a  molecular envelope (radius of 300~AU and velocity of 5~\kms) onto an intermediate- to high-mass protostar. 
We estimate an infall rate of $6 \times 10^{-4}~n_{8}$~\ms~yr$^{-1}$ , where $n_{8}$ is the ambient volume density in units of  10$^{8}$~cm$^{-3}$ (required for maser excitation). 
In addition, our measurements of H$_2$O maser (and radio continuum)  emission  identify a collimated bipolar molecular outflow (and ionized jet) from MM--1. The evidence of simultaneous accretion and outflow at small spatial scales, makes
AFGL 5142 an extremely compelling target for high-angular resolution studies of high-mass star formation. 
}

   \keywords{
Masers -- Star formation -- Stars: circumstellar matter  -- ISM: jets and outflows -- ISM: individual (AFGL 5142) 
}

                \maketitle

%

\section{Introduction}
Theoretical models of star formation predict that protostars form from the gravitational collapse of molecular cloud cores  and mass accretes onto the protostar either  by direct accretion of low angular momentum material or via a disk 
\citep[e.g.,][]{Shu87}. 
Hence, observational signatures of infalling envelopes in early stages 
are essential to progress in the understanding of the mass-accretion process and constrain theoretical models of star formation.
To date, bulk infall motions have been detected mostly by observing inverse P-Cygni profiles of molecular lines 
along the line-of-sight (l.o.s.) to the infalling gas against a bright background (dust or H\,{\small II} emission),
 in both low-mass \citep[e.g.,][]{Lee09}
 and high-mass \citep[e.g.,][]{Beltran11} protostars. 
Nonetheless, discerning the signature of infall in embedded protostars is inherently difficult.
First, significant infall motions are confined only to the innermost regions of the core, and the small spatial scales involved make a direct measurement difficult. 
Second, other phenomena can mimic infall. Line asymmetries may be associated with competing processes such as rotation and outflows, or even arise from superposition of intervening clouds along the l.o.s. 
Confusion is more severe in the case of high-mass protostars,  which are on average more distant ($>$1 kpc) and form embedded in rich protosclusters.
Evidence of {\it global} infall has been found only in a handful of high-mass  star forming regions
 and at large radii ($>$1000~AU) from the protostar(s) forming at the center \citep{Zhang97,Sollins05,Beltran06,Beltran11}.
Besides, {\it even for the few known bona fide cases of infall}, ambiguity arises from comparison of observations at different angular resolutions. 
On the one hand, molecular {\it thermal} transitions (e.g., CH$_3$CN, NH$_3$, CO), observed with present connected-element interferometers, trace infall/rotation at radii of several thousands of AU \citep{Beltran06,Beltran11}. 
On the other hand, molecular {\it masers} (e.g.,  H$_2$O), observed with Very Long Baseline Interferometry (VLBI), clearly trace expansion at much smaller radii of hundreds of AU \citep{Goddi05,Mosca07}. 
Thus, in order to definitely establish whether the rotating and infalling material at large scales eventually accretes onto individual protostars, infall motions should be measured in the proximity ($\ll$1000 AU) to the protostar.

In this letter, we report  a convincing signature of infall of  a circumstellar molecular envelope with a radius of only 300~AU in the well-known high-mass star-forming region AFGL 5142 (1.8~kpc; \citealt{Sne88}). Observations with the Submillimeter Array (SMA) at 1.3~mm identified a high-mass protocluster containing five dusty mm cores and three CO outflows, within a few arcseconds \citep{Zhang07}. The focus of this letter is the mm core MM--1, which shows hot-core chemistry, exhibits radio continuum emission from ionized gas, and powers strong water and methanol masers \citep{Goddi07}, indicating that it is likely the most massive object in the cluster.  

New multi-epoch VLBI observations of \met~masers enabled us to measure the 3-D velocity field of molecular gas, providing the most direct and most unbiased measurement (yet obtained) of infall 
 onto an intermediate- to high-mass protostar.
We also discuss the kinematics of \wat~masers and the physical properties of radio continuum emission with the aim of characterizing  the outflow structure, 
and hence providing a complete picture of star formation
 on scales $<$400 AU, usually not accessible by observations yet relevant to test accretion models.

\begin{table}
\caption{Summary of Observations toward AFGL 5142 MM--1.}             
\label{obs}      
\centering                        
\begin{tabular}{ccccccc} 
\hline\hline                 
\noalign{\smallskip}
Emission & $\nu_{\rm rest}$ & Array & Date & HPBW & P.A. \\
 & (GHz) &   & (yy/mm/dd) &  (mas $\times$ mas) & $(^{\circ})$\\
\noalign{\smallskip}
\hline                        
\noalign{\smallskip}
\met   & 6.669    &  EVN  & 04/11/04 &  $5.9 \times 5.4$ & 20 \\
\met   & 6.669    &  EVN  & 07/03/16 &  $6.3 \times 5.8$ & --88 \\
\met   & 6.669    &  EVN  & 09/03/12 &  $5.6 \times 4.3$ &48  \\
H$_2$O   & 22.235   &  VLBA  & 03/10/16 &  $0.7 \times 0.4$ & 0\\
H$_2$O   & 22.235   &  VLBA  & 03/11/22 &  $0.7 \times 0.4$ & 0\\
H$_2$O   & 22.235   &  VLBA  & 04/01/01 &  $0.7 \times 0.4$ & 0\\
 H$_2$O   & 22.235   &  VLBA  & 04/02/08 &  $0.7 \times 0.4$ & 0\\
cont   & 8.4     &  VLA-A & 03/06/15 &  $188 \times 129$ & 39 \\
cont   & 22.2    &  VLA-B & 05/02/28 &  $248 \times 226$ & 0 \\
\noalign{\smallskip}
\hline   
\end{tabular}
\end{table}
\begin{figure}
\centering
\includegraphics[angle= -90,trim=0.47cm 0 0 0 ,width=0.48\textwidth]{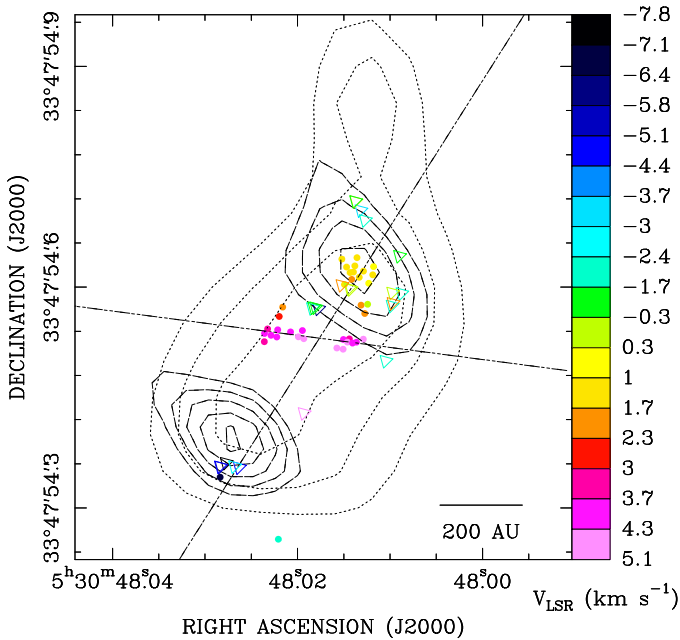}
\caption{ 
Molecular masers and radio continuum in AFGL 5142  MM--1. Positions and l.o.s. velocities of H$_2$O masers (observed with the VLBA) are indicated by triangles and CH$_3$OH masers (observed with the EVN) are indicated by circles. Colors code l.o.s. velocities, according to the wedge to the right (the systemic velocity, --1.1 \kms, is in green). The contour maps show the continuum emissions (observed with the VLA) at 22~GHz (dotted contours, representing 3, 4, and 5 times the 46 $\mu$Jy~beam$^{-1}$ rms noise level) and 8.4 GHz (dashed contours,  representing 3, 4, 5, 6 and 7 times the 23~$\mu$Jy~beam$^{-1}$ rms noise level).
The two small-dashed lines show the linear fits to the positions of all water masers (P.A. --32\degree) and of only the red methanol masers (P.A. 82\degree).
}
\label{positions}
\end{figure}


\section{Observations and results}
For six years, our team has launched an observational campaign of different molecular masers and  radio continuum emission in AFGL 5142 using the European VLBI Network (EVN), 
 the Very Long Baseline Array (VLBA), and the Very Large Array (VLA).
Table~\ref{obs} summarizes the observations.  
We observed with the EVN  the  $5_1\to6_0 A^+$ transition of  \met\ at 6.669~GHz
 at three distinct epochs, over the years 2004--2009. Data of the first epoch were reported in \citet{Goddi07} and we remand the reader to that article for a full description of the EVN observing setup and data reduction. We discuss here for the first time the subsequent epochs and proper motion measurements of \met\ masers.
 The \(6_{16}-5_{23}\) transition of \wat\ at 22.235~GHz was observed with the VLBA  at four distinct epochs over the years 2003--2004; \citet{Goddi06} discuss in detail  the 3-D kinematics of \wat\ masers. 
Here, we also report new images of the continuum emission from VLA archival data at 22~GHz \citep{Goddi06} and 8.4~GHz  \citep{Zhang07}. 
Published maps  used {\it natural} weighting of the {\it uv}-data and had an angular resolution $\sim$0\pas3--0\pas4.  
The new VLA maps at both frequencies were produced with {\it robust} weighting, improving the angular resolution to 0\pas16 (8.4~GHz) and 0\pas24 (22~GHz).

Figure~1 shows positions and l.o.s. velocities of the 22 GHz water and  the 6.7 GHz methanol  masers (first epoch of observation)overplotted on the contour maps of the continuum emission at 8.4~GHz and 22~GHz.
The 22~GHz continuum emission appears elongated northwest-southeast (NW-SE), while the 8.4~GHz emission (with higher-angular resolution) is resolved in two components separated by 300 mas (or 540~AU) on the plane of the sky. 
\wat~masers are concentrated in two clusters, associated with the two components of the 8.4~GHz continuum: the one towards the SE has l.o.s. velocities blue-shifted  with respect to the systemic velocity of the region (--1.1~\kms; \citealt{Zhang07}); the other one located to the NW has red-shifted l.o.s. velocities.
\met~masers are distributed across a similar area as the \wat~masers, and consist of three clusters. A cluster of only two features with blue-shifted l.o.s. velocities is associated with the SE continuum peak (not discussed here); a second cluster with red-shifted l.o.s. velocities is associated with  the NW continuum peak ("yellow" features); a third cluster with the most red-shifted l.o.s. velocities has an intermediate position between the two 8.4~GHz continuum peaks 
 ("red"  features).
The dashed lines in Fig.~\ref{positions} provide the best (least-square) fit  line of positions of all the water masers (position angle or P.A. of --32\degree) and of only the  red  methanol masers (P.A. of 82\degree), respectively.  

\begin{figure}
\includegraphics[angle= -90,width=0.5\textwidth]{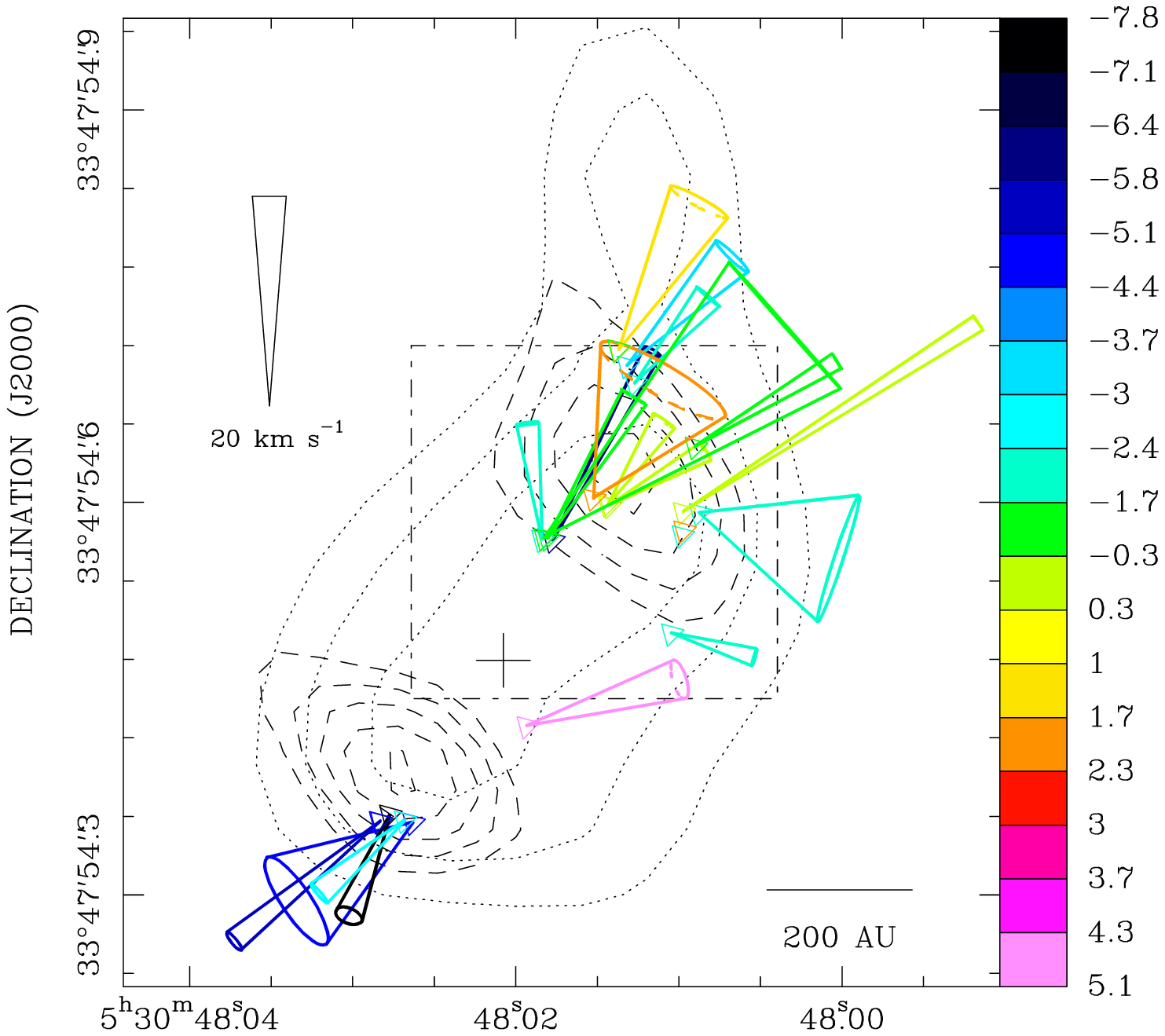}
\includegraphics[angle= -90,width=0.5\textwidth]{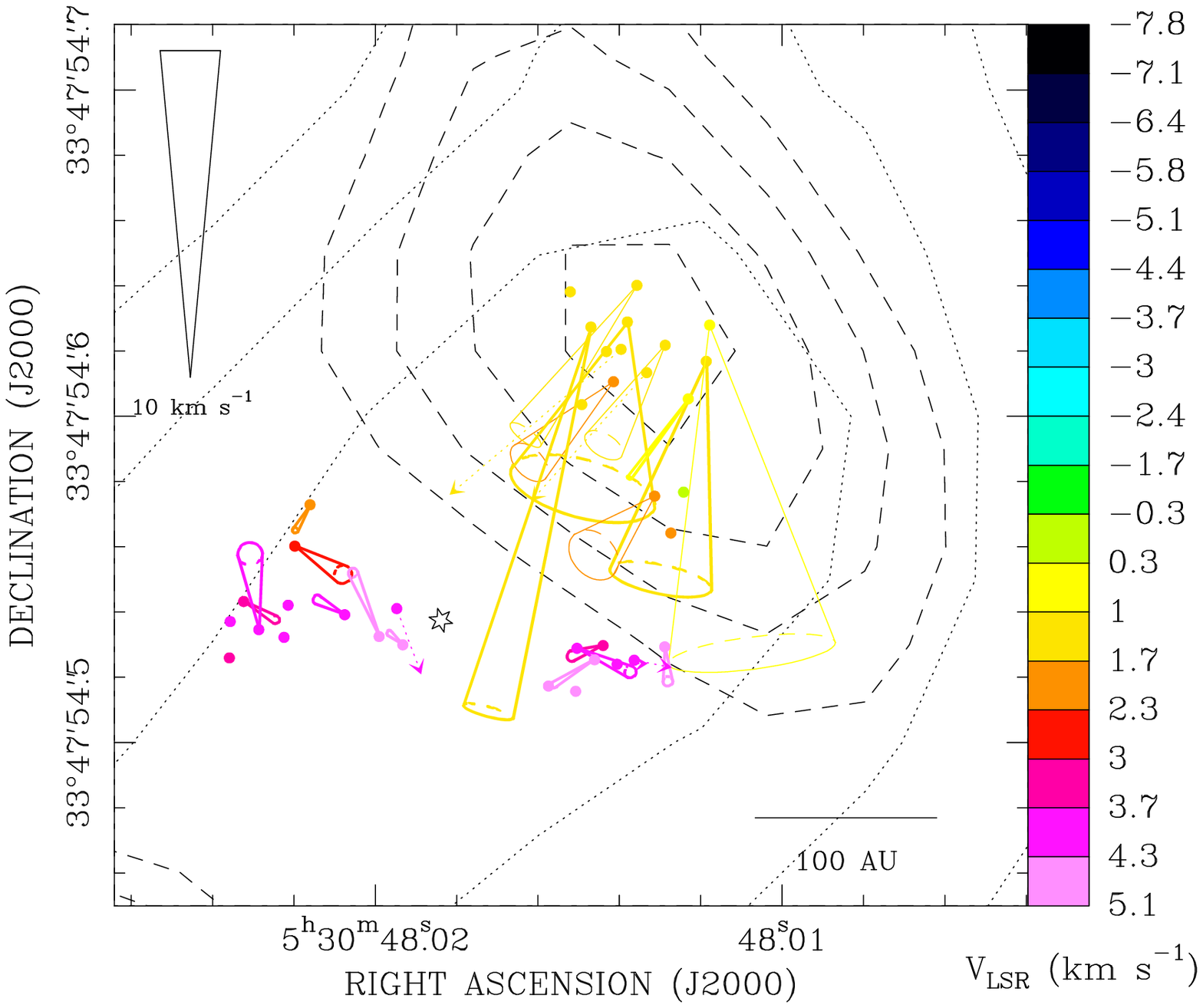}
\caption{ 
Proper motions of H$_2$O ({\it upper panel}) and CH$_3$OH ({\it lower panel})  masers  in AFGL 5142 MM--1, as measured relative to their centers of motion, 
independently calculated  for water ({\it cross}) and methanol ({\it star}) masers. The  rectangle in the upper panel shows the area plotted in the lower panel.
The cones indicate orientation and uncertainties of measured proper motions (the amplitude scale is given in each panel) and 
colors denote l.o.s. velocities. Contour maps show the VLA 22~GHz (dotted line) and 8.4 GHz (dashed line) continuum emissions.
%
%
The black star in the lower panel identifies the putative location of the protostar.
}
\label{prmot}
\end{figure}

Figure~\ref{prmot} shows the proper motions 
of \wat\ masers (upper panel)  
and \met\ masers (lower panel).
Since measurements of {\it absolute} proper motions are affected by the combined uncertainty 
of the Solar motion and Galactic rotation curve (up to 15 \kms; \citealt{Reid09}), we prefer to base our analysis on {\it relative} velocities  for both water and methanol masers. 
This requires to adopt a suitable reference system, ideally centered on the protostar.   
For water,  we calculated the geometric mean of positions at individual epochs for all the masers persistent over four epochs  (hereafter "center of motion") and we referred the proper motions to this point. 
This method is equivalent to subtracting to all masers the average proper motion of the selected features.
The relative proper motions indicate that the two clusters are moving away from each other along a NW-SE direction, with velocities $\sim$20 \kms.
Although showing a similar pattern, the relative proper motions identify a more collimated bipolar flow than the absolute proper motions (compare with  Fig. 2 in \citealt{Goddi07}). 
We can now use positions and 3-D velocities of water masers to define geometric parameters of the outflow from  MM--1 (for simplicity, supposed of conical shape): P.A.  of the sky-projected axis, inclination angle of the axis with the plane of the sky, and opening angle. 
The median P.A. of relative proper motions (--47\degree) is similar to the P.A. of the best-fit line of water maser positions (--32\degree) and we can assume an intermediate value (--40\degree) for the P.A. of the sky-projected outflow axis.  The semi-opening angle of the outflow (25\degree) is estimated from the maximum angle between the proper motion P.A. and the outflow axis.
To estimate the inclination angle, we reasoned as follows. 
 Based on the ratio of l.o.s. velocities and proper motions, water masers appear to move mainly on the plane of the sky. 
If they arise from the surface of a cone, then the outflow inclination angle   can be approximated with its semi-opening angle (i.e., 25\degree). 

Since the amplitude of methanol proper motions  is much lower than that of water (with mean values $\sim$3 \kms~and $\sim$15 \kms, respectively), 
the choice of a suitable reference for methanol velocities is more critical. 
The red  masers have  {\it internal} proper motions
 (calculated using the center of motion of only red masers; 1--2 \kms) 
much smaller than the internal proper motions of the yellow masers (relative to the center
of motion of only yellow features; 1--10 \kms);  
 the opposite is true for their l.o.s. velocities (3--6 \kms~vs. 0--2 \kms).
Since red masers  are projected in the plane of the sky closer to the putative protostellar position 
(see below) and move mostly along the l.o.s., 
 we assume that their  average proper motion gives an estimate of the protostar proper motion.   
Hence, we calculated the center of motion of methanol using  a sample of only red features with a stable spatial and spectral structure.  
 The resulting proper motions, shown in Fig.~\ref{prmot}b, should represent the sky-projected velocities 
as roughly measured  by an observer comoving with the star.\footnote{Our analysis provides the kinematic of methanol masers {\it relative} to the protostar, 
but not the peculiar protostellar velocity, which can be much smaller than the uncertainty 
 on the measured {\it absolute} proper motions of methanol masers ($\sim$10--20 \kms).} 
The yellow masers have proper motions with larger amplitudes and directed towards the centroid of the red masers, 
which instead move mostly along the l.o.s.; both aspects indicate infall towards the protostar. 
This is consistent with methanol masers having red-shifted l.o.s. velocities and being in the foreground of the 22 GHz continuum emission (optically thick at 6.7 GHz).
 The center of motion calculated from the red cluster of methanol masers provides our best estimate of the location of the protostar 
 (marked with a star in Fig.~\ref{prmot}b). 
Three lines of arguments support this hypothesis: 
1) the red masers are distributed in a symmetric and elongated structure roughly perpendicular to the outflow axis traced by water masers; 
2)  the two best-fit lines of water and red methanol masers intersect approximately in correspondence of the methanol center of motion, 
close to the centroid of the 22~GHz continuum (Fig.~\ref{positions});
3) the model employed to interpret the methanol kinematics gives a similar best-fit location of the protostar 
(within uncertainties $\sim$20 mas; see Sect. \ref{dis_infall}).

%


\section{Discussion}
\label{discussion}
 Our VLBI observations show that methanol and water masers, despite being excited at similar radii from MM--1,  trace  different kinematics: outflow, infall, and (perhaps) rotation, which we discuss separately in the following sub-sections.

\subsection{Collimated outflow from MM--1}
We identified  a collimated bipolar outflow,  at radii 140 to 400~AU from the driving protostar, traced by  radio continuum emission in its ionized component and by water masers in its molecular component.
Before this work, the radio continuum could be interpreted as either a hyper-compact (HC) H\,{\small II} region or an ionized jet \citep{Goddi06}. The high-angular resolution maps in Fig.~\ref{prmot} reveal that the two radio continuum  peaks are associated with the two expanding clusters of water masers.  This evidence favours the interpretation of the radio continuum as an ionized jet 
and suggests that water masers may originate in shocks produced by the interaction of the jet with the surrounding molecular environment. 
For optically thin emission, $F \, d^2 = 10^{3.5} \, (\Omega/4\pi) \, \dot{P_j}$ \citep{Sanna10a}, where $F$ is the measured continuum flux in mJy, $\dot{P_j}$ is the jet momentum rate in \mbox{\ms~yr$^{-1}$~\kms}, $\Omega$ is the jet solid angle in sr, and $d$ is the source distance in kpc.  Using a flux density of 0.24~mJy (0.17~mJy) measured at 8.4~GHz for the NW (SE) component and a distance of 1.8~kpc, we derive:   $ \dot{P_j} = 2.5 (1.7) \times 10^{-4} \, (\Omega/4\pi)^{-1}$~\mbox{\ms~yr$^{-1}$~\kms}. 

Based on our measurements of positions and 3-D velocities of water masers and under reasonable assumptions for gas densities, 
we can estimate the mass-loss rate ($\dot{M}$) and the momentum rate ($\dot{P_o}$=$\dot{M} V$) of the molecular outflow. 
For a conical flow with solid angle $\Omega$, the mass transferred across a shell  of thickness $dr$ in a time interval $dt$  at a distance $R$ from the protostar, is
 $\dot{M} = 4\pi R^2 n_{H_2} m_{H_2} V \, (\Omega/4\pi)$ (where $V$=$dr/dt$).  
We then derive:  $ \dot{M} = 1.5 \times 10^{-4} \, V_{10} \, R_{100}^{2} \, (\Omega/4\pi) \, n_{8}$ \mbox{\ms~yr$^{-1}$} and $ \dot{P_o} =  1.5 \times 10^{-3} \, V_{10}^{2} \, R_{100}^{2} \, (\Omega/4\pi) \, n_{8}$ \mbox{\ms~yr$^{-1}$~\kms}, where  $V_{10}$  is the average maser velocity in units of 10 \kms, $R_{100}$  is the average distance of water masers in units of 100 AU, and $n_{8}$ is the ambient volume density in units of  10$^{8}$ cm$^{-3}$ (required for water maser excitation; e.g., \citealt{Kau96}). 
The average (sky-projected) distance of water masers from the protostar is 290~AU and their average velocity is 15 \kms,  
which give 
$ \dot{M} = 1.9  \times 10^{-3}  \, (\Omega/4\pi)\; n_{8}$ \mbox{\ms~yr$^{-1}$} and $ \dot{P_o} = 2.8  \times 10^{-2} \, (\Omega/4\pi) \; n_{8}$ \mbox{\ms~yr$^{-1}$~\kms}.  
If the same outflow is responsible for exciting the radio continuum and water masers, the solid angle $\Omega$
can be determined by requiring that the momentum rate in the
maser outflow equals that from the continuum emission.
We derive  $\Omega$=1--1.2 sr, corresponding to an outflow semi-opening angle of 32--36\degree, intermediate between the angle determined by water masers (25\degree) and the angle subtended by the two  radio continuum components (90\degree). 
For a semi-opening angle of 34\degree, we derive $\dot{M} = 1.6  \times 10^{-4} \; n_{8}$ \mbox{\ms~yr$^{-1}$} and $ \dot{P} = 2.4  \times 10^{-3} \; n_{8}$ \ms yr$^{-1}$ \kms, indicating  strong outflow activity from MM--1.

 On larger scales, CO (2--1) line emission revealed a multiple system of outflows, oriented north-south, northeast-southwest, and NW-SE \citep{Zhang07}. The poor angular resolution of the SMA (1--3\arcsec) precluded a clear identification of the driving sources of outflows.
Our higher-angular resolution measurements allow us to identify unambiguously  MM--1 as the driving source of the NW-SE oriented outflow, traced by \wat~masers and radio continuum at radii $<$400AU and by CO emission at radii $>$2000~AU.
\citet{Zhang07} derived a momentum rate of $2\times10^{-3}$~\msyr~\kms, 
which is in good agreement with the momentum rate of the \wat~maser outflow.
We conclude that the ionized jet observed in the radio continuum emission and the molecular outflow probed by  water masers, can account  for the acceleration of the large-scale CO outflow driven by  MM--1.


\begin{figure}
\centering
\includegraphics[angle= -90,width=0.5\textwidth]{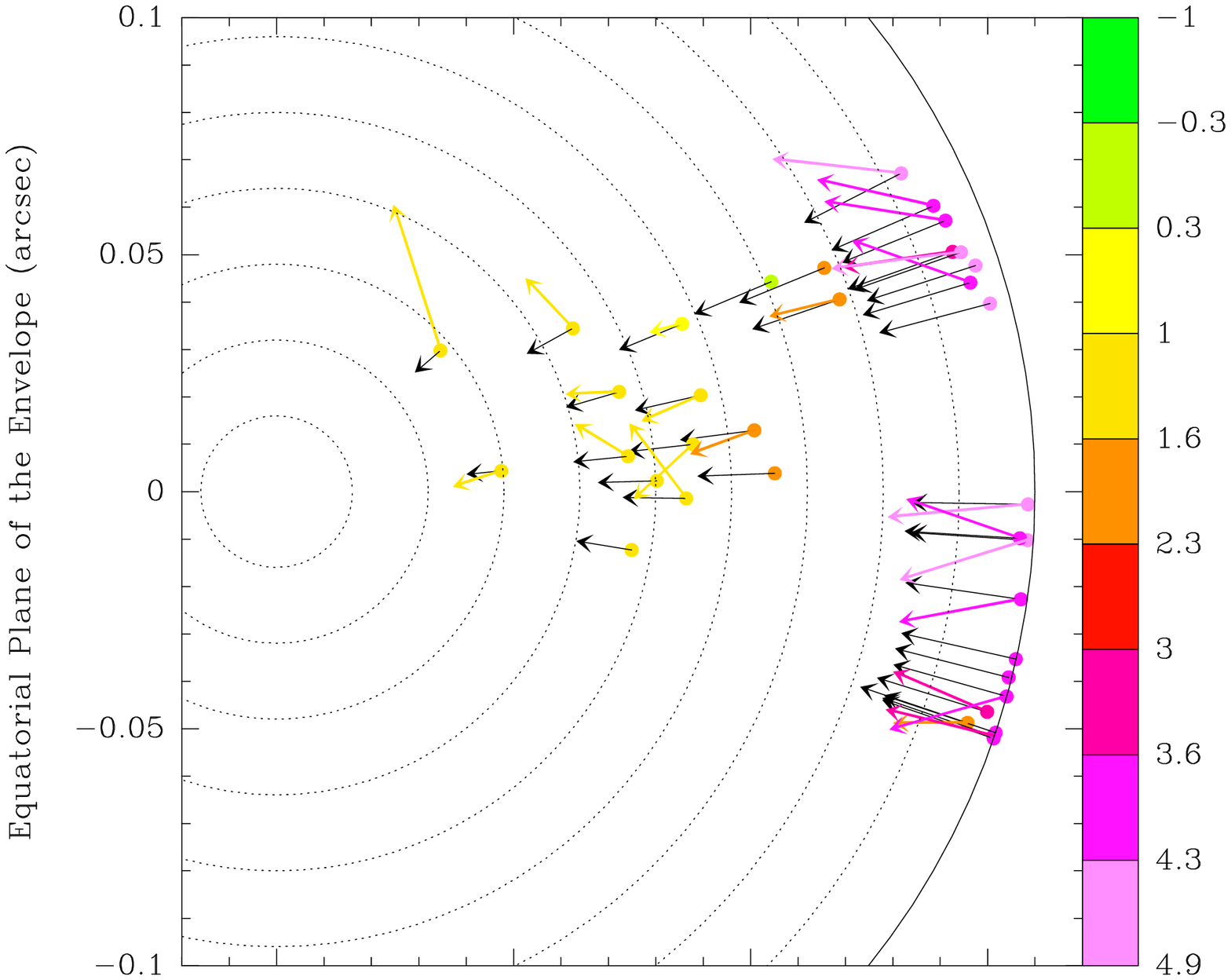}
\includegraphics[angle= -90,trim=0.26cm 0 0cm 0,width=0.5\textwidth]{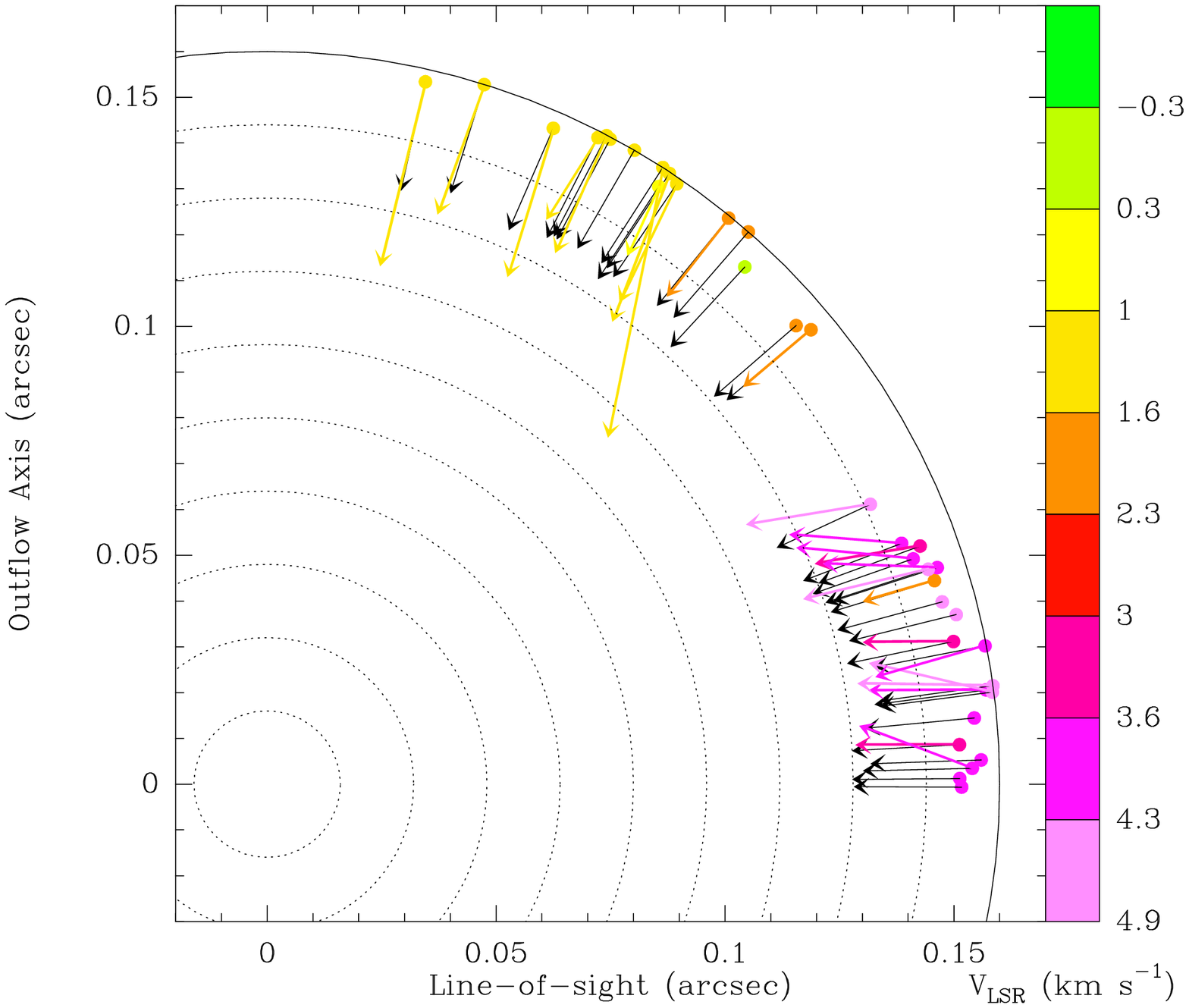}
\caption{ Measured ({\it color arrows}) and best-fit model  ({\it black arrows}) 3-D velocity vectors  of CH$_3$OH  masers.  We show projections onto the equatorial plane of the envelope ({\it upper panel}) and a plane containing the l.o.s. and the outflow axis ({\it lower panel}). 
We assume an outflow axis lying on the plane of the sky 
 (we obtain qualitatively the same picture by assuming an outflow inclination angle of 25\degree). 
 The  modeled infalling envelope has a radius of  0\pas16 (or 290 AU) and an infall velocity of 5~\kms. The dotted lines indicate concentric circles at steps of 10\% of the radius (0\pas016) around the protostar at the (0,0) position.}
\label{infall}
\end{figure}


\subsection{Infall of gas onto MM--1 from a molecular envelope}
\label{dis_infall}
Infall  onto  MM--1 was first claimed by \citet{Goddi07}, based solely on positions and l.o.s. velocities of methanol masers. The measurement of proper motions 
 has provided the missing kinematic observable to confirm the infall hypothesis.  

\begin{figure}
\centering
\includegraphics[angle=-90,trim=0 0 0cm 0,width=0.4\textwidth]{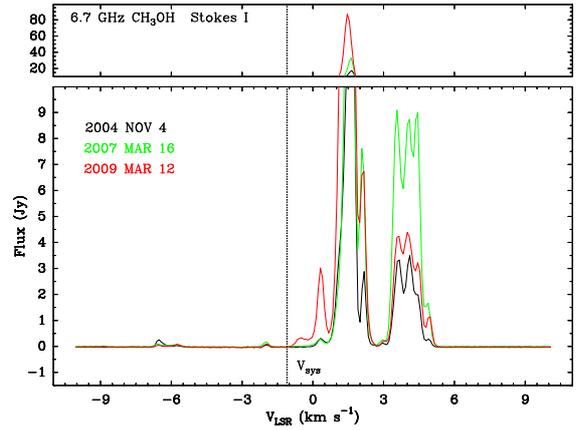}
\caption{ 
Total-power spectra of CH$_3$OH emission toward MM--1.
 The upper panel shows a zoom of the methanol maser peak. 
 The vertical dotted line represents the systemic velocity 
(--1.1~\kms; \citealt{Zhang07}). 
Note that the emission in the velocity range 0--3~\kms\ (yellow masers) is always stronger than emission in the range 3--6~\kms\ (red masers).
}
\label{spectra}
\end{figure}

We adopt a simple spherical model to derive main parameters of infall: 
 $ {\bf V}({\bf R}) = \sqrt{\frac{2 \ G \ M}{R^{3}}} \ {\bf R} $, 
where ${\bf V}$  is the velocity field, ${\bf R}$ is the position vector  from the protostar, $M$ is the total gas mass within a sphere of radius $R$ centered on the protostar. 
The model has four free parameters: the sky-projected coordinates ($\alpha$ and $\delta$) of the protostar (the sphere center), the sphere radius $R$, and $M$, sampled in the range [--0\pas1, 0\pas1], [--0\pas1, 0\pas1], [0\pas1, 0\pas4], and [1, 25~\ms], respectively. The best-fit parameters are found by using fine steps (1\%) of the parameter ranges and minimizing the $\chi^{2}$ given by the sum of the squared differences between the 3-D  velocity vectors of the model and the methanol data: 
 $R=290\pm70$~AU = 0\pas16 $\pm$ 0\pas03 and  $M=4\pm1$~\ms, corresponding to $V_{\rm inf}=5\pm1$~\kms. 
 For each fit parameter, the  error corresponds to a variation from the best fit value for which the  $\chi^{2}$ value increases by $\sim$10\%. 
Figure~\ref{infall} shows the measured and best-fit model 3-D velocity vectors of methanol masers
projected onto the plane containing the protostar and perpendicular to the outflow axis (the equatorial plane; upper panel) and the plane containing the l.o.s. and the outflow axis (lower panel). The best-fit position of the protostar (0,0) is coincident (within uncertainties$\sim$0\pas02) with the  methanol center of motion. 
Most of the measured 3-D velocities are well reproduced by the infall model.
The two projections show also that methanol masers  do not sample the whole spherical envelope but concentrate into preferred areas.
The red features sample a relatively small solid angle about the l.o.s. to the protostar: their
distance from the protostar (0$\farcs$15) is about three times 
their maximum offset along the equatorial plane (0$\farcs$05).
The yellow features are found closer to the pole of the sphere.
This behaviour may be explained considering that molecular masers are preferably observed in regions where 
either strong background emission can be amplified or
long, velocity-coherent, amplification paths along the l.o.s. occur. 
The yellow features lie on top of the NW peak of the radio continuum, which naturally provides the background radiation being amplified. 
For the red features, projected close the putative protostellar position, possible background sources (undetected by the VLA) are: a HC~H\,{\small II} region of 0$\farcs$05~radius; 
 the base of an ionized wind within 0$\farcs$05~from the protostar. 
The elongation of red features could be explained with the larger densities expected  in the equatorial plane, providing longer  amplification path for the maser radiation. 
The amplification of the radio continuum background is consistent with the yellow masers (located on top of the continuum peak) being much stronger than the red masers (see the total-power spectra of the 6.7 GHz emission in Fig.~\ref{spectra}).


We can now use the best-fit parameters for the shell radius  and  infall velocity derived from our model, to estimate a mass infall rate, $\dot{M}_{\rm inf}=4 \pi R^2  n_{H_2} m_{H_2} V$,  and an infall momentum rate, $\dot{P}_{\rm inf}=\dot{M}_{\rm inf} V_{\rm inf}$. For  $R=290$~AU and $V_{\rm inf}=5$~\kms, we obtain: 
$\dot{M}_{\rm inf}  = 6 \times 10^{-4}~n_{8}$~\msyr\ and $\dot{P}_{\rm inf}=3 \times 10^{-3}~n_{8}$~\ms~km~s$^{-1}$~yr$^{-1}$, where $n_{8}$ is the gas volume density in units of  $10^{8}$~cm$^{-3}$ (required for methanol maser excitation; \citealt{Cragg05}).  
On the one hand, the high value estimated for the infall rate might indicate that the protostar is in an active accretion stage. On the other hand, a ratio of 0.27 for the mass-loss rate to the infall rate  indicates that  the outflow can efficiently remove mass (and angular momentum) from the system, as expected from magneto-centrifugal ejection. 
We caution that  the mass rates estimated above are accurate only to within 1--2
orders of magnitude, owing to uncertainty in the  density required for maser excitation.
Nevertheless, our measurements show that the two maser species used to derive the mass-loss rate (\wat) and the mass infall rate (\met) are distributed across a similar area, making plausible the hypothesis that they are excited in molecular gas with similar densities. Hence, the dependence of the outflow/infall rate ratio on gas density should be less critical. 

We estimated 4~\ms~for the total gas mass enclosed in the infalling envelope, which is consistent with the value estimated from dust continuum emission (3~\ms; \citealt{Zhang07})\footnote{Note that \citet{Goddi07}  assumed a wrong systemic velocity of --4.4 \kms\ (derived from low-angular resolution data), which resulted in a 2$\times$ infall velocity and a 4$\times$ central mass (using only l.o.s velocities).}. 
Several lines of evidence suggest however the presence of a protostar with mass $\gg$4~\ms~at the center of the envelope: the high luminosity of the region ($>10^4$~\ls), the large mass of the CO outflow (3~\ms), the presence of methanol masers (detected only in high-mass protostars), and the powerful outflow activity evidenced by strong water maser and radio continuum emission. 
In fact, the simple approach adopted here considers only gravity but likely non-gravitational forces  can influence  gas dynamics. 
\citet{Girart09} found recently that the gravitational collapse of a massive  molecular core in G31.41+0.31 is controlled by magnetic fields, which appear to be effective in removing angular momentum and  slowing down  gas infall, as expected from magnetic braking \citep{Galli06}. 
Recently, \citet{Matthews10}  measured SiO maser proper motions in the disk-wind system associated with Orion Source I and estimated  a dynamical mass of 7~\ms\ for the central protostar, much lower than the estimate from the dynamics of interacting protostars in the region \citep[20~\ms;][]{Goddi11}. 
\citet{Matthews10} suggested that magnetic fields may be important in driving the gas dynamics around Source I  
and \citet{Goddi11} argued that, assuming magnetic support in the disk, a significant fraction of the mass of the central object can be "hidden" in the Keplerian profile of maser  velocities,  leading to an underestimate of the mass.
Similar mechanisms could be at work in  AFGL 5142  MM--1.
Besides magnetic fields, any realistic calculation of the infall of a circumstellar envelope should also include the effects of rotation. 
As the collapse proceeds, assuming the angular momentum is conserved, rotation should lead eventually to the formation of a centrifugally supported disk around the accreting protostar (see next Sect.), which 
would  contribute to further underestimate the actual mass of the central protostar. 

\subsection{Is there a rotating disk perpendicular to the outflow?}
Our methanol maser measurements provide hints of the presence of a rotating disk. 
The red masers are elongated across 300 AU in the plane of the sky roughly perpendicular to the jet axis. 
Some yellow features closer to the jet axis show also a large residual  between modeled and measured velocities (Fig.~\ref{infall}), which could be interpreted in terms of anticlockwise rotation about the jet.
However, the good fit of the 3-D velocities of methanol masers obtained with a spherical infall model, 
implies that infall dominates the velocity field   at  300~AU from the protostar.
Hence, we expect the disk to be compact  (radius $<$ 300 AU).
The disk hypothesis is also supported by recent  interferometric imaging of complex molecules, 
which reveals a structure of size $\sim$600 AU, elongated east-west and centered on the protostellar position derived here, within astrometric uncertainties \citep{Palau11}.


The simultaneous presence of a collimated outflow, an infalling envelope, and (possibly) a rotating disk, makes AFGL 5142 an extremely compelling target for high-angular resolution studies of  accretion  in massive star formation. 
\begin{acknowledgements}
  A.S. acknowledges financial support by the ERC Advanced Investigator Grant GLOSTAR (247078).
\end{acknowledgements}

\bibliography{biblio}  
\bibliographystyle{aa}
%
%

\end{document}